\documentclass[a4paper,10pt]{article}
\usepackage{stmaryrd}
\usepackage{amsfonts}
\usepackage{bbm}
\usepackage{amscd}
\usepackage{mathrsfs}
\usepackage{latexsym,amssymb,amsmath,amscd,amscd,amsthm,amsxtra,xypic}
\usepackage[dvips]{graphicx}
\usepackage[utf8]{inputenc}
\usepackage[T1]{fontenc}
\usepackage{lmodern}
\usepackage{amssymb}
\usepackage[all]{xy}
\usepackage{nicefrac,mathtools,enumitem}
\usepackage{microtype}

\newtheorem{thm}{Theorem}[section]
\newtheorem{lem}[thm]{Lemma}

\newtheorem{pro}[thm]{Proposition}

\newtheorem{rmk}[thm]{Remark}
\newtheorem{defi}[thm]{Definition}

\newcommand{\lon }{\,\rightarrow\,}
\newcommand{\be }{\begin{equation}}
\newcommand{\ee }{\end{equation}}

\newcommand{\pf}{\noindent{\bf Proof.}\ }

\newcommand{\frkg}{\mathfrak g}
\newcommand{\frkh}{\mathfrak h}

\newcommand{\frkk}{\mathfrak k}

\newcommand{\frkG}{\mathfrak G}

\def\qed{\hfill ~\vrule height6pt width6pt depth0pt}

\newcommand{\br}[1]{   [ \cdot,    \cdot  ]_\frkg   }

\newcommand{\Id}{\rm{Id}}

\newcommand{\dM}{\mathrm{d}}

\newcommand{\Der}{\mathrm{Der}}
\newcommand{\Inn}{\mathrm{Inn}}

\newcommand{\gl}{\mathfrak {gl}}

\newcommand{\ad}{\mathrm{ad}}

\begin{document}
\title{
{Representations of hom-Lie algebras
\thanks
 {
Research partially supported by  NSF of China (11026046) and
Doctoral Fund. of MEC (20100061120096).
 }
} }
\author{Yunhe Sheng  \\
Department of Mathematics, Jilin University,\\
 Changchun 130012, Jilin, China
\\\vspace{3mm}
email: shengyh@jlu.edu.cn }

\date{}


\maketitle
\begin{abstract}
In this paper, we study representations of hom-Lie algebras. In
particular, the adjoint representation and the trivial
representation of hom-Lie algebras are studied in detail.
Derivations,  deformations, central extensions and derivation
extensions of hom-Lie algebras are also studied as an application.

\end{abstract}

\section{Introduction}

The notion of hom-Lie algebras was introduced by Hartwig, Larsson,
and Silvestrov in \cite{HLS} as part of a study of deformations of
the Witt and the Virasoro algebras. In a hom-Lie algebra, the Jacobi
identity is twisted by a linear map, called the hom-Jacobi identity.
Some $q$-deformations of the Witt and the Virasoro algebras have the
structure of a hom-Lie algebra \cite{HLS}. Because of close relation
to discrete and deformed vector fields and differential calculus
\cite{HLS,LD1,LD2}, hom-Lie algebras are widely studied recently
\cite{AEM,BM,MS1,MS2,Yao1,Yao2,Yao3}.

 In nowadays mathematics, much
of the research on certain algebraic object is to study its
representation theory. The representation theory of an algebraic
object reveals some of its profound structures hidden underneath. A
good example is that the structure of a complex semi-simple Lie
algebra is much revealed via its representation theory.  However,
according to the author's knowledge, the representation theory of
hom-Lie algebras is not so well developed. Donald Yau  defined the
hom-L-modules for hom-Lie algebras and studied the corresponding
homology in \cite{Yao2}. Makhlouf and Silvestrov  studied the formal
deformations of hom-Lie algebras in \cite{MS1}, where only
$1$-hom-cochains, $2$-hom-cochains, $1$-coboundary operator and
$2$-coboundary operator are defined without the help of any
representation.

The purpose of this paper is to fulfill this gap, i.e. to define
representations of hom-Lie algebras and corresponding hom-cochain
complexes. In particular, we obtain the adjoint representation and
the trivial representation of hom-Lie algebras. We give a detail
study on the cohomologies associated to the adjoint representation
and the trivial representation. A very interesting phenomenon is
that associated to a (regular) hom-Lie algebra
$(\frkg,\br,,\alpha)$, there is a series of adjoint representations
$\ad_s$, which we call $\alpha^s$-adjoint representation, defined by
$$
\ad_su(v)=[\alpha^s(u),v]_\frkg.
$$
As an application,  we study derivations,  deformations, central
extensions and derivation extensions of hom-Lie algebras in detail.
In \cite{AEM}, the authors also construct a cochain complex
associated to a multiplicative hom-Lie algebra independently and
study the one-parameter formal deformation of multiplicative hom-Lie
algebras. It turns out that the cochain complex therein corresponds
to the $\alpha^{-1}$-adjoint representation in this paper.

The paper is organized as follows. In Section 2 after giving the
definition of hom-Lie algebras, we show that the direct sum of two
hom-Lie algebras is still a hom-Lie algebra. A linear map between
hom-Lie algebras is a morphism if and only if its graph is a hom-Lie
sub-algebra. In Section 3 we study derivations of multiplicative
hom-Lie algebras. For any nonnegative integer $k$, we define
$\alpha^k$-derivations of the multiplicative hom-Lie algebra
$(\frkg,\br ,,\alpha)$. Considering the direct sum of the space of
$\alpha^k$-derivations, we prove that it is a Lie algebra
(Proposition \ref{pro:derivationLiea}). In particular, any
$\alpha$-derivation gives rise to a derivation extension of the
multiplicative hom-Lie algebra $(\frkg,\br ,,\alpha)$ (Theorem
\ref{thm:Dextension}). In Section 4 we give the definition of
representations of multiplicative hom-Lie algebras and the
corresponding coboundary operators. We show that one can obtain the
semidirect product multiplicative hom-Lie algebra $(\frkg\oplus
V,[\cdot,\cdot]_{\rho_A},\alpha+A)$ associated to any representation
$\rho_{A}$ on $V$ of the multiplicative hom-Lie algebra $(\frkg,\br
,,\alpha)$ (Proposition \ref{pro:semidirectproduct}). In Section 5
we study the trivial representations of multiplicative hom-Lie
algebras. We show that
 central extensions of a multiplicative hom-Lie algebra are controlled by the second cohomology with
 coefficients in the trivial representation (Theorem
 \ref{thm:centralextension}). In Section 6
we study the adjoint representation of a regular hom-Lie algebra
$(\frkg,\br ,,\alpha)$. For any integer $s$, we define the
$\alpha^s$-adjoint representation. We show that a $1$-cocycle
associated to the $\alpha^s$-adjoint representation is exactly an
$\alpha^{s+1}$-derivation of the regular hom-Lie algebra $(\frkg,\br
,,\alpha)$ (Proposition \ref{pro:derivation}). In Subsection 6.1 we
mainly study the $\alpha^{-1}$-adjoint representation. We show that
similar to the case of Lie algebras, any  deformation of a regular
hom-Lie algebra is controlled by its second cohomology with
coefficients in the $\alpha^{-1}$-adjoint representation. We also
give the definition of hom-Nijenhuis operators of regular hom-Lie
algebras. We show that the deformation generated by a hom-Nijenhuis
operator is trivial. In Subsection 6.2 we analyze the cohomology of
hom-Lie algebras with coefficients in the $\alpha$-adjoint
representation. \vspace{3mm}

{\bf Acknowledgements:} We would like to give our warmest thanks to
Zhangju Liu and Abdenacer Makhlouf for very helpful comments and
also give our warmest thanks to Chenchang Zhu for the help during we
stayed in Courant Research Center, G$\ddot{\rm{o}}$ttingen, where a
part of work was done. We also give our special thanks to referees
for many helpful suggestions.

\section{Hom-Lie algebras}

In this paper, we follow a slightly more general definition of
Hom-Lie algebras from \cite{MS2,BM}.

\begin{defi}\begin{itemize}\item[\rm(1)]
  A hom-Lie algebra is a triple $(\frkg,\br ,,\alpha)$ consisting of a
  vector space $\frkg$, a skew-symmetric bilinear map (bracket) $\br,:\wedge^2\frkg\longrightarrow
  \frkg$ and a linear map $\alpha:\frkg\lon\frkg$ satisfying the following hom-Jacobi
  identity:
  \begin{equation}
    [\alpha(u),[v,w]_\frkg]_\frkg+[\alpha(v),[w,u]_\frkg]_\frkg+[\alpha(w),[u,v]_\frkg]_\frkg=0;
  \end{equation}

 \item[\rm(2)] A hom-Lie algebra is called
a multiplicative hom-Lie algebra if $\alpha$ is an algebraic
morphism, i.e.  for any $u,v\in\frkg$, we have $\alpha([u,v]_\frkg)
= [\alpha(u), \alpha(v)]_\frkg;$

\item[\rm(3)]A hom-Lie algebra is called a regular hom-Lie algebra if $\alpha$ is
an algebra automorphism;

 \item[\rm(4)] A sub-vector space $\frkh\subset\frkg$ is a hom-Lie sub-algebra of $(\frkg,\br ,,\alpha)$ if
 $\alpha(\frkh)\subset\frkh$ and
  $\frkh$ is closed under the bracket operation $\br,$, i.e.
  $$
[u,u^\prime]_\frkg\in\frkh,\quad\forall~u,u^\prime \in\frkh.
  $$
  \end{itemize}
\end{defi}

Consider the direct sum of two hom-Lie algebras, we have

\begin{pro}
Given two hom-Lie algebras $(\frkg,\br ,,\alpha)$ and
$(\frkk,[\cdot,\cdot]_\frkk,\beta)$, there is a hom-Lie algebra
$(\frkg\oplus\frkk,[\cdot,\cdot]_{\frkg\oplus\frkk},\alpha+\beta)$,
where the skew-symmetric bilinear map
$[\cdot,\cdot]_{\frkg\oplus\frkk}:\wedge^2(\frkg\oplus\frkk)\longrightarrow
  \frkg\oplus\frkk$ is given by
  $$
 [(u_1,v_1),(u_2,v_2)]_{\frkg\oplus\frkk}=([u_1,u_2]_\frkg,[v_1,v_2]_\frkk),\quad\forall~u_1,u_2\in\frkg,~v_1,v_2\in\frkk,
  $$
  and the linear map
  $(\alpha+\beta):\frkg\oplus\frkk\longrightarrow \frkg\oplus\frkk$ is given by
$$
  (\alpha+\beta)(u,v)=(\alpha(u),\beta(v)),\quad\forall~u\in\frkg,~v\in\frkk.
$$
\end{pro}

A morphism of hom-Lie algebras $\phi:(\frkg,\br
,,\alpha)\longrightarrow(\frkk,[\cdot,\cdot]_\frkk,\beta)$ is a
linear map $\phi:\frkg\longrightarrow\frkk$ such that
\begin{eqnarray}
\label{eqn:phimorphism1}\phi[u,v]_\frkg&=&[\phi(u),\phi(v)]_\frkk,\quad\forall~u,v\in\frkg,\\\label{eqn:phimorphism2}
\phi\circ\alpha&=&\beta\circ\phi.
\end{eqnarray}

Denote by $\frkG_\phi\subset\frkg\oplus\frkk$  the graph of a linear
map $\phi:\frkg\longrightarrow\frkk$.

\begin{pro}
 A linear map $\phi:(\frkg,\br
,,\alpha)\longrightarrow(\frkk,[\cdot,\cdot]_\frkk,\beta)$ is a
morphism of hom-Lie algebras if and only if the graph
$\frkG_\phi\subset\frkg\oplus\frkk$ is a hom-Lie sub-algebra of
$(\frkg\oplus\frkk,[\cdot,\cdot]_{\frkg\oplus\frkk},\alpha+\beta)$.
\end{pro}
\pf Let $\phi:(\frkg,\br
,,\alpha)\longrightarrow(\frkk,[\cdot,\cdot]_\frkk,\beta)$ be a
morphism of hom-Lie algebras, then for any $u,v\in\frkg$, we have
\begin{eqnarray*}
  [(u,\phi(u)),(v,\phi(v))]_{\frkg\oplus\frkk}=([u,v]_\frkg,[\phi(u),\phi(v)]_\frkk)=([u,v]_\frkg,\phi[u,v]_\frkg).
\end{eqnarray*}
Thus the graph $\frkG_\phi$ is closed under the bracket operation
$[\cdot,\cdot]_{\frkg\oplus\frkk}$. Furthermore, by
\eqref{eqn:phimorphism2}, we have
$$
(\alpha+\beta)(u,\phi(u))=(\alpha(u),\beta\circ\phi(u))=(\alpha(u),\phi\circ\alpha(u)),
$$
which implies that $(\alpha+\beta)(\frkG_\phi)\subset\frkG_\phi$.
Thus $\frkG_\phi$ is a hom-Lie sub-algebra of
$(\frkg\oplus\frkk,[\cdot,\cdot]_{\frkg\oplus\frkk},\alpha+\beta)$.

Conversely, if the graph $\frkG_\phi\subset\frkg\oplus\frkk$ is a
hom-Lie sub-algebra of
$(\frkg\oplus\frkk,[\cdot,\cdot]_{\frkg\oplus\frkk},\alpha+\beta)$,
then we have
\begin{eqnarray*}
  [(u,\phi(u)),(v,\phi(v))]_{\frkg\oplus\frkk}=([u,v]_\frkg,[\phi(u),\phi(v)]_\frkk)\in\frkG_\phi,
\end{eqnarray*}
which implies that
$$
[\phi(u),\phi(v)]_\frkk=\phi[u,v]_\frkg.
$$
Furthermore, $(\alpha+\beta)(\frkG_\phi)\subset\frkG_\phi$ yields
that
$$
(\alpha+\beta)(u,\phi(u))=(\alpha(u),\beta\circ\phi(u))\in\frkG_\phi,
$$
which is equivalent to the condition
$\beta\circ\phi(u)=\phi\circ\alpha(u)$, i.e.
$\beta\circ\phi=\phi\circ\alpha$. Therefore, $\phi$ is a morphism of
hom-Lie algebras. \qed

\section{Derivations of hom-Lie algebras}
Let $(\frkg,\br ,,\alpha)$ be a multiplicative hom-Lie algebra. For
any nonnegative integer $k$,  denote by $\alpha^k$ the $k$-times
composition of $\alpha$, i.e.
$$
\alpha^k=\alpha\circ\cdots\circ\alpha ~(\mbox{$k$-times}).
$$
In particular, $~\alpha^0=\Id$ and $\alpha^1=\alpha$. If $(\frkg,\br
,,\alpha)$ is a regular hom-Lie algebra, we denote by $\alpha^{-k}$
the $k$-times composition of $\alpha^{-1}$, the inverse of $\alpha$.

\begin{defi}\label{defi:derivation}
For any nonnegative integer $k$, a linear map $D:\frkg\lon\frkg$ is
called an $\alpha^k$-derivation of the multiplicative hom-Lie
algebra $(\frkg,\br ,,\alpha)$, if
\begin{equation}\label{derivationC1}
[D,\alpha]=0,\quad \mbox{i.e.} \quad D\circ\alpha=\alpha\circ D,
\end{equation}
and
\begin{equation}\label{derivationC2}
  D[u,v]_\frkg=[D(u),\alpha^k(v)]_\frkg+[\alpha^k(u),D(v)]_\frkg,\quad\forall~u,v\in\frkg.
\end{equation}
For a regular hom-Lie algebra, $\alpha^{-k}$-derivations can be
defined similarly.
\end{defi}
\begin{rmk}
  In \cite{MS1}, the cohomologies in degree-$1$ and degree-$2$ of hom-Lie algebras are
  defined. a linear map $D:\frkg\lon\frkg$ is an $\alpha^0$-derivation is
  equivalent to that $D$ is a $1$-cocycle. In Section $6$ we will
  introduce $\alpha^s$-adjoint representations of hom-Lie algebras.
  We will show that $D$ is an $\alpha^{s+1}$-derivation if and only
  if $D$ is a $1$-cocycle associated to the $\alpha^s$-adjoint
  representation.
\end{rmk}

Denote by $\Der_{\alpha^k}(\frkg)$ the set of $\alpha^k$-derivations
of the multiplicative hom-Lie algebra $(\frkg,\br ,,\alpha)$. For
any $u\in\frkg$ satisfying $\alpha(u)=u$, define
$D_k(u):\frkg\lon\frkg$ by
$$
D_k(u)(v)=[\alpha^k(v),u]_\frkg,\quad\forall~v\in\frkg.
$$
Then $D_k(u)$ is an $\alpha^{k+1}$-derivation, which we call an {\bf
inner} $\alpha^{k+1}$-derivation. In fact, we have
$$
D_k(u)(\alpha(v))=[\alpha^{k+1}(v),u]_\frkg=\alpha([\alpha^{k}(v),u]_\frkg)=\alpha\circ
D_k(u)(v),
$$
which implies that \eqref{derivationC1} in Definition
\ref{defi:derivation} is satisfied. On the other hand, we have
\begin{eqnarray*}
  D_k(u)([v,w]_\frkg)&=&[\alpha^k([v,w]_\frkg),u]_\frkg=[[\alpha^k(v),\alpha^k(w)]_\frkg,\alpha(u)]_\frkg\\
  &=&[\alpha^{k+1}(v),[\alpha^k(w),u]_\frkg]_\frkg+[[\alpha^k(v),u]_\frkg,\alpha^{k+1}(w)]_\frkg\\
  &=&[\alpha^{k+1}(v),D_k(u)(w)]_\frkg+[D_k(u)(v),\alpha^{k+1}(w)]_\frkg.
\end{eqnarray*}
Therefore, $D_k(u)$ is an $\alpha^{k+1}$-derivation. Denote by
$\Inn_{\alpha^k}(\frkg)$ the set of inner $\alpha^{k}$-derivations,
i.e.
\begin{equation}
 \Inn_{\alpha^k}(\frkg)=\{[\alpha^{k-1}(\cdot),u]_\frkg|~u\in\frkg,~\alpha(u)=u\}.
\end{equation}

 For any
$D\in\Der_{\alpha^k}(\frkg) $ and
$D^\prime\in\Der_{\alpha^s}(\frkg), $
 define their commutator $[D,D^\prime]$ as usual:
\begin{equation}\label{eqn:commutator}
[D,D^\prime]=D\circ D^\prime-D^\prime\circ D.
\end{equation}
\begin{lem}\label{lem:derivation}
 For
any $D\in\Der_{\alpha^k}(\frkg) $ and
$D^\prime\in\Der_{\alpha^s}(\frkg), $  we have
$$[D,D^\prime]\in\Der_{\alpha^{k+s}}(\frkg).$$
\end{lem}
\pf For any $u,v\in\frkg$, we have
\begin{eqnarray*}
[D,D^\prime]([u,v]_\frkg)&=&D\circ
D^\prime([u,v]_\frkg)-D^\prime\circ
D([u,v]_\frkg)\\
&=&D([D^\prime(u),\alpha^s(v)]_\frkg+[\alpha^s(u),D^\prime(v)]_\frkg)-D^\prime([D(u),\alpha^k(v)]_\frkg+[\alpha^k(u),D(v)]_\frkg)\\
&=&[D\circ D^\prime(u),\alpha^{k+s}(v)]_\frkg+[\alpha^k\circ
D^\prime(u),D\circ\alpha^s(v)]_\frkg\\
&&+[D\circ\alpha^s(u),\alpha^k\circ
D^\prime(v)]_\frkg+[\alpha^{k+s}(u),D\circ D^\prime(v)]_\frkg\\
&&-[D^\prime\circ D(u),\alpha^{k+s}(v)]_\frkg-[ \alpha^s\circ
D(u),D^\prime\circ\alpha^{k}(v)]_\frkg\\
&&-[D^\prime\circ\alpha^k(u),\alpha^s\circ
D(v)]_\frkg-[\alpha^{k+s}(u),D^\prime \circ D(v)]_\frkg.
\end{eqnarray*}
Since $D$ and $D^\prime$ satisfy
$$
D\circ \alpha=\alpha\circ D,\quad D^\prime\circ \alpha=\alpha\circ
D^\prime,
$$
we have
$$
\alpha^k\circ D^\prime=D^\prime\circ \alpha^k,\quad
D\circ\alpha^s=\alpha^s\circ D.
$$
Therefore, we have
\begin{eqnarray*}
[D,D^\prime]([u,v]_\frkg)&=&[\alpha^{k+s}(u),[D,D^\prime](v)]_\frkg+[[D,
D^\prime](u),\alpha^{k+s}(v)]_\frkg.
\end{eqnarray*}
Furthermore, it is straightforward to see that
$$
[D,D^\prime]\circ\alpha=D\circ D^\prime\circ\alpha-D^\prime\circ
D\circ\alpha=\alpha\circ D\circ D^\prime-\alpha\circ D^\prime\circ
D=\alpha\circ [D,D^\prime],
$$
which yields that
$[D,D^\prime]\in\Der_{\alpha^{k+s}}(\frkg)$.\qed\vspace{3mm}


Denote by
\begin{equation}
  \Der(\frkg)=\bigoplus_{k\geq0}\Der_{\alpha^k}(\frkg).
\end{equation}

By Lemma \ref{lem:derivation}, obviously we have
\begin{pro}\label{pro:derivationLiea}
  With the above notations, $\Der(\frkg)$ is a Lie algebra, in which
  the Lie bracket is given by \eqref{eqn:commutator}.
\end{pro}
\begin{rmk}
  Similarly, we can obtain a Lie algebra
  $\bigoplus_{k}\Der_{\alpha^k}(\frkg)$, where $k$ is any integer,
  for a regular hom-Lie algebra.
\end{rmk}

At the end of this section, we consider the derivation extension of
the multiplicative hom-Lie algebra
$(\frkg,[\cdot,\cdot]_\frkg,\alpha)$ and give an application of the
$\alpha$-derivation $\Der_{\alpha}(\frkg)$.

For any linear map $D:\frkg\lon\frkg$, consider the vector space
$\frkg\oplus \mathbb RD$. Define a skew-symmetric bilinear bracket
operation $[\cdot,\cdot]_D$ on $\frkg\oplus \mathbb RD$ by
$$
[u,v]_D=[u,v]_\frkg,\quad [D,u]_D=-[u,D]_D=D(u),\quad\forall
~u,v\in\frkg.
$$
Define a linear map $\alpha_D:\frkg\oplus \mathbb RD\lon \frkg\oplus
\mathbb RD$ by $\alpha_D(u,D)=(\alpha(u),D)$, i.e.
$$
\alpha_D=\left(\begin{array}{ll}\alpha&0\\0&1\end{array}\right).
$$
\begin{thm}\label{thm:Dextension}
With the above notations, $(\frkg\oplus \mathbb
RD,[\cdot,\cdot]_D,\alpha_D)$ is a multiplicative hom-Lie algebra if
and only if $D$ is an $\alpha$-derivation of the multiplicative
hom-Lie algebra $(\frkg,[\cdot,\cdot]_\frkg,\alpha)$.
\end{thm}
\pf For any $u,v\in \frkg,~m,n\in\mathbb R$, we have
\begin{eqnarray*}
 \alpha_D [(u,mD),(v,nD)]_D&=& \alpha_D ([u,v]_\frkg+mD(v)-nD(u))\\
 &=&\alpha([u,v]_\frkg)+m\alpha\circ D(v)-n\alpha\circ D(u),
\end{eqnarray*}
and
\begin{eqnarray*}
  [\alpha_D(u,mD),\alpha_D(v,nD)]_D&=&
  [(\alpha(u),mD),(\alpha(v),nD)]_D\\
  &=&[\alpha(u),\alpha(v)]_\frkg+mD\circ\alpha(v)-nD\circ\alpha(u).
\end{eqnarray*}
Since $\alpha$ is an algebra morphism, thus $\alpha_D $ is an
algebra morphism if and only if
$$
D\circ\alpha=\alpha\circ D.
$$
 On the other hand, we have
\begin{eqnarray*}
 &&
 [\alpha_D(D),[u,v]_D]_D+[\alpha_D(u),[v,D]_D]_D+[\alpha_D(v),[D,u]_D]_D\\
 &=&D([u,v]_\frkg)-[\alpha(u),D(v)]_\frkg-[D(u),\alpha(v)]_\frkg.
\end{eqnarray*}
Therefore, it is obvious that the hom-Jacobi identity is satisfied
if and only if
$$
D([u,v]_\frkg)-[\alpha(u),D(v)]_\frkg-[D(u),\alpha(v)]_\frkg=0.
$$
Thus $(\frkg\oplus \mathbb RD,[\cdot,\cdot]_D,\alpha_D)$ is a
multiplicative hom-Lie algebra if and only if $D$ is an
$\alpha$-derivation of $(\frkg,[\cdot,\cdot]_\frkg,\alpha)$.
\qed\vspace{3mm}

The author studied the derivation extension of 3D-dimensional Lie
algebras and therefore gave the classification of 4D-dimensional Lie
algebras using Poisson method in \cite{sheng}.

\section{Representations of hom-Lie algebras}
In this section we study representations of multiplicative hom-Lie
algebras and give the corresponding coboundary operators. We also
prove that one can form  semidirect product multiplicative hom-Lie
algebras when given representations of multiplicative hom-Lie
algebras. Please see \cite{CE,Jacobson} for more details about Lie
algebras and their cohomologies. Let $(\frkg,\br,,\alpha)$ be a
multiplicative hom-Lie algebra and $V$ be an arbitrary vector space.
Let $A\in\gl(V)$ be an arbitrary linear transformation from $V$ to
$V$.
\begin{defi}
  A representation of the multiplicative hom-Lie algebra $(\frkg,\br,,\alpha)$ on
  the vector space $V$ with respect to $A\in\gl(V)$ is a linear map
  $\rho_A:\frkg\longrightarrow \gl(V)$, such that for any
  $u,v\in\frkg$, the following equalities are satisfied:
  \begin{eqnarray}
 \label{representation1} \rho_A(\alpha(u))\circ A&=&A\circ \rho_A(u);\\\label{representation2}
    \rho_A([u,v]_\frkg)\circ
    A&=&\rho_A(\alpha(u))\circ\rho_A(v)-\rho_A(\alpha(v))\circ\rho_A(u).
  \end{eqnarray}
\end{defi}
The set of $k$-cochains on $\frkg$ with values in $V$, which we
denote by $C^k(\frkg;V)$, is the set of  skew-symmetric $k$-linear
maps from $\frkg\times\cdots\times\frkg$ $(k$-times$)$ to $V$:
$$
C^k(\frkg;V)\triangleq\{f:\wedge^k\frkg\longrightarrow V ~\mbox{is a
linear map}\}.
$$

A $k$-hom-cochain on $\frkg$ with values in $V$ is defined to be a
$k$-cochain $f\in C^k(\frkg;V)$ such that it is compatible with
$\alpha$ and $A$ in the sense that $A\circ f=f\circ \alpha$, i.e.
$$
A(f(u_1,\cdots,u_k))=f(\alpha(u_1),\cdots,\alpha(u_k)).
$$
Denote by $ C^k_{\alpha,A}(\frkg;V)$ the set of $k$-hom-cochains:
$$
C^k_{\alpha,A}(\frkg;V)\triangleq\{f\in C^k(\frkg;V)|~A\circ
f=f\circ \alpha\}.
$$

Define $\dM_{\rho_A}:C^k_{\alpha,A}(\frkg;V)\longrightarrow
C^{k+1}(\frkg;V)$ by setting
\begin{eqnarray*}
  \dM_{\rho_A} f(u_1,\cdots,u_{k+1})&=&\sum_{i=1}^{k+1}(-1)^{i+1}\rho_A(\alpha^{k}(u_i))(f(u_1,\cdots,\widehat{u_i},\cdots,u_{k+1}))\\
  &&+\sum_{i<j}(-1)^{i+j}f([u_i,u_j]_\frkg,\alpha(u_1)\cdots,\widehat{u_i},\cdots,\widehat{u_j},\cdots,\alpha(u_{k+1})).
\end{eqnarray*}
\begin{lem}
  With the above notations, for any $f\in C^k_{\alpha,A}(\frkg;V)$, we have
  $$(\dM_{\rho_A} f)\circ\alpha=A\circ\dM_{\rho_A}
  f.$$ Thus we obtain a well-defined map
  $$
\dM_{\rho_A}:C^k_{\alpha,A}(\frkg;V)\longrightarrow
C^{k+1}_{\alpha,A}(\frkg;V).
  $$
\end{lem}
\pf The conclusion follows from the facts that $f\circ\alpha=A\circ
f$  and $\alpha$ is an algebra morphism. More precisely, we have
\begin{eqnarray*} &&\dM_{\rho_A} f(\alpha(u_1),\cdots,\alpha(u_{k+1}))\\
&=&\sum_{i=1}^{k+1}(-1)^{i+1}\rho_A(\alpha^{k+1}(u_i))(f(\alpha(u_1),\cdots,\widehat{u_i},\cdots,\alpha(u_{k+1)}))\\
  &&+\sum_{i<j}(-1)^{i+j}f([\alpha(u_i),\alpha(u_j)]_\frkg,\alpha^2(u_1)\cdots,\widehat{u_i},\cdots,\widehat{u_j},\cdots,\alpha^2(u_{k+1}))\\
  &=&\sum_{i=1}^{k+1}(-1)^{i+1}\rho_A(\alpha^{k+1}(u_i))\circ A\circ f(u_1,\cdots,\widehat{u_i},\cdots,u_{k+1})\\
  &&+\sum_{i<j}(-1)^{i+j}f(\alpha[u_i,u_j]_\frkg,\alpha^2(u_1)\cdots,\widehat{u_i},\cdots,\widehat{u_j},\cdots,\alpha^2(u_{k+1}))\\
  &=&\sum_{i=1}^{k+1}(-1)^{i+1}A\circ\rho_A(\alpha^{k}(u_i))(f(u_1,\cdots,\widehat{u_i},\cdots,u_{k+1}))\\
  &&+\sum_{i<j}(-1)^{i+j}A\circ
  f([u_i,u_j]_\frkg,\alpha(u_1)\cdots,\widehat{u_i},\cdots,\widehat{u_j},\cdots,\alpha(u_{k+1}))\\
  &=&A\circ  \dM_{\rho_A} f(u_1,\cdots,u_{k+1}),
\end{eqnarray*}
which completes the proof. \qed

\begin{rmk}
The condition $A\circ f=f\circ \alpha$ in the definition of
$k$-hom-cochains is necessary as we will see in the proof of
$\dM_{\rho_A}^2=0$. When considering $\frkg$ represents on itself,
i.e. the adjoint representation, this condition reduces to
$\alpha\circ f=f\circ\alpha$. Note that in Definition 5.2 in
\cite{MS1}, this condition is omitted.
\end{rmk}

\begin{pro}\label{pro:coboundary}
The map $\dM_{\rho_A}$ is a coboundary operator, i.e.
$\dM_{\rho_A}^2=0$.
\end{pro}

The proof of this proposition is ``standard", which is given in
Appendix.\vspace{3mm}

Associated to the representation $\rho_A$, we obtain the complex
$(C^k_{\alpha,A}(\frkg;V),\dM_{\rho_A})$. Denote the set of closed
$k$-hom-cochains by $Z^k(\frkg;\rho_A)$ and the set of exact
$k$-hom-cochains by $B^k(\frkg;\rho_A)$. Denote the corresponding
cohomology by
$$
H^k(\frkg;\rho_A)=Z^k(\frkg;\rho_A)/B^k(\frkg;\rho_A).
$$

In the case of Lie algebras, we can form semidirect products when
given representations. Similarly, we have
\begin{pro}\label{pro:semidirectproduct}
  Given a representation $\rho_A$ of the multiplicative hom-Lie algebra $(\frkg,\br,,\alpha)$ on
  the vector space $V$ with respect to $A\in\gl(V)$. Define a
 skew-symmetric bilinear bracket operation $[\cdot,\cdot]_{\rho_A}:\wedge^2(\frkg\oplus V)\longrightarrow\frkg\oplus
 V$ by
 \begin{equation}
   [(u,X),(v,Y)]_{\rho_A}=([u,v]_\frkg,\rho_A(u)(Y)-\rho_A(v)(X)).
 \end{equation}
 Define $\alpha+A:\frkg\oplus V\longrightarrow \frkg\oplus V$ by
 $$
(\alpha+A)(u,X)=(\alpha(u),AX).
 $$
 Then $(\frkg\oplus V,[\cdot,\cdot]_{\rho_A},\alpha+A)$ is a multiplicative hom-Lie
 algebra, which we call the semidirect product of the multiplicative hom-Lie algebra
 $(\frkg,\br,,\alpha)$ and $V$.
\end{pro}
\pf First we show that $\alpha+A$ is an algebra morphism. On one
hand, we have
\begin{eqnarray*}
  (\alpha+A)[(u,X),(v,Y)]_{\rho_A}&=&
  (\alpha+A)([u,v]_\frkg,\rho_A(u)(Y)-\rho_A(v)(X))\\
  &=&(\alpha([u,v]_\frkg),A\circ\rho_A(u)(Y)-A\circ\rho_A(v)(X)).
\end{eqnarray*}
On the other hand, we have
\begin{eqnarray*}
  [(\alpha+A)(u,X),(\alpha+A)(v,Y)]_{\rho_A}&=&
  [(\alpha(u),AX),(\alpha(v),AY)]_{\rho_A}\\
  &=&([\alpha(u),\alpha(v)]_\frkg,\rho_A(\alpha(u))(AY)-\rho_A(\alpha(v))(AX)).
\end{eqnarray*}
Since $\alpha$ is an algebra morphism, $\rho_A$ and $A$ satisfy
\eqref{representation1}, it follows that $\alpha+A$ is an algebra
morphism with respect to the bracket $[\cdot,\cdot]_{\rho_A}$. It is
not hard to deduce that
\begin{eqnarray*}
 && [(\alpha+A)(u,X),[(v,Y),(w,Z)]_{\rho_A}]_{\rho_A}+c.p.((u,X),(v,Y),(w,Z))\\&=&
  [\alpha(u),[v,w]_\frkg]_\frkg+c.p.(u,v,w)\\
  &&+\rho_A(\alpha(u))\circ\rho_A(v)(Z)-\rho_A(\alpha(v))\circ\rho_A(u)(Z)-\rho_A([u,v]_\frkg)\circ
  A(Z)\\&&-\rho_A(\alpha(u))\circ\rho_A(w)(Y)-\rho_A([w,u]_\frkg)\circ
  A(Y)+\rho_A(\alpha(w))\circ\rho_A(u)(Y)\\
  &&-\rho_A([v,w]_\frkg)\circ
  A(X)+\rho_A(\alpha(v))\circ\rho_A(w)(X)-\rho_A(\alpha(w))\circ\rho_A(v)(X),
\end{eqnarray*}
where $c.p.(a,b,c)$ means the cyclic permutations of $a,b,c$. By
\eqref{representation2}, the hom-Jacobi identity is satisfied. Thus,
$(\frkg\oplus V,[\cdot,\cdot]_{\rho_A},\alpha+A)$
 is a  multiplicative  hom-Lie algebra. \qed

\section{The trivial representation of hom-Lie algebras}
In this section, we study the trivial representation of
multiplicative hom-Lie algebras. As an application, we show that the
central extension of a multiplicative hom-Lie algebra
$(\frkg,\br,,\alpha)$ is controlled by the second cohomology of
$\frkg$ with coefficients in the trivial representation.

Now let $V=\mathbb R$, then we have $\gl(V)=\mathbb R$. Any
$A\in\gl(V)$ is exactly a real number, which we use the notation
$r$. Let $\rho:\frkg\longrightarrow\gl(V)=\mathbb R$ be the zero
map. Obviously, $\rho$ is a representation of the multiplicative
hom-Lie algebra $(\frkg,\br,,\alpha)$ with respect to any $r\in
\mathbb R$. We will always assume that $r=1$. We call this
representation the {\bf trivial representation} of the
multiplicative hom-Lie algebra $(\frkg,\br,,\alpha)$.

Associated to the trivial representation, the set of $k$-cochains on
$\frkg$, which we denote by $C^k(\frkg)$, is the set of
skew-symmetric $k$-linear maps from $\frkg\times\cdots\times\frkg$
to $\mathbb R$, i.e. $ C^k(\frkg)=\wedge^k\frkg^*.$ The set of
$k$-hom-cochains is given by
$$
 C^k_{\alpha}(\frkg)=\{f\in\wedge^k\frkg^*|f\circ\alpha=f\}.
$$
The corresponding coboundary operator $\dM_T:
C^k_{\alpha}(\frkg)\longrightarrow  C^{k+1}_{\alpha}(\frkg)$ is
given by
$$
\dM_T
f(u_1,\cdots,u_{k+1})=\sum_{i<j}(-1)^{i+j}f([u_i,u_j]_\frkg,\alpha(u_1),\cdots,\widehat{u_i},\cdots,\widehat{u_j},\cdots,\alpha(u_{k+1})).
$$
Denote by $Z^k_\alpha(\frkg)$ and $B^k_\alpha(\frkg)$ the
corresponding closed $k$-hom-cochains and exact $k$-hom-cochains
respectively. Denote the resulting cohomology by $H^k(\frkg)$.

\begin{pro}
  With the above notations, associated to the trivial representation of
  the multiplicative hom-Lie algebra $(\frkg,\br,,\alpha)$, we have
\begin{eqnarray*}
  H^0(\frkg)&=&\mathbb R;\\
  H^1(\frkg)&=&\{f\in C^1_{\alpha}(\frkg)|~f|_{[\frkg,\frkg]_\frkg}=0\}.
\end{eqnarray*}
\end{pro}
\pf Obviously,  any $s\in\mathbb R$ is a $0$-hom-cochain. By the
definition of coboundary operator $\dM_T$, we have $\dM_Ts=0$. Thus
we have $ H^0(\frkg)=\mathbb R.$

For any $f\in C^1_{\alpha}(\frkg)$, we have
$$
\dM_Tf(u,v)=-f([u,v]_\frkg).
$$
Therefore, $f$ is closed if and only if
$f|_{[\frkg,\frkg]_\frkg}=0$. The conclusion follows from the fact
that there is no exact $1$-hom-cochain. \qed\vspace{2mm}

In the following we consider central extensions of the
multiplicative hom-Lie algebra $(\frkg,\br,,\alpha)$. We will see
that it is controlled by the second cohomology $H^2(\frkg)$.

Obviously, $(\mathbb R,0,1)$ is an abelian multiplicative hom-Lie
algebra with the trivial bracket and the identity morphism. Let
$\theta\in C^2_\alpha(\frkg)$, we consider the direct sum
$\frkh=\frkg\oplus \mathbb R$ with the following bracket
\begin{equation}
[(u,s),(v,t)]_\theta=([u,v]_\frkg,\theta(u,v)),\quad
\forall~u,v\in\frkg,~s,t\in\mathbb R.
\end{equation}
Define $\alpha_\frkh:\frkh\longrightarrow\frkh$ by
$\alpha_\frkh(u,s)=(\alpha(u),s)$, i.e.
$$
\alpha_\frkh\triangleq\left(\begin{array}{ll}\alpha&0\\0&1\end{array}\right).
$$

\begin{thm}\label{thm:centralextension}
  The triple $(\frkh,[\cdot,\cdot]_\theta,\alpha_\frkh)$ is a multiplicative hom-Lie
  algebra if and only if $\theta\in C^2_\alpha(\frkg)$ is a $1$-cocycle associated to the trivial representation, i.e.
  $$\dM_T\theta=0.$$
\end{thm}
 We call the multiplicative hom-Lie algebra
 $(\frkh,[\cdot,\cdot]_\theta,\alpha_\frkh)$ the {\em central
 extension}
 of  $(\frkg,\br,,\alpha)$ by the abelian hom-Lie algebra $(\mathbb R,0,1)$.

 \pf The fact that $\alpha_\eta$ is an algebra morphism with respect to the bracket
 $[\cdot,\cdot]_\theta$ follows from the fact that
 $\theta\circ\alpha=\theta$. More precisely, we have
$$
\alpha_\eta[(u,s),(v,t)]_\theta=(\alpha[u,v]_\frkg,\theta(u,v)).
$$
On the other hand, we have
$$
[\alpha_\eta(u,s),\alpha_\eta(v,t)]_\theta=[(\alpha(u),s),(\alpha(v),t)]_\theta=([\alpha(u),\alpha(v)]_\frkg,\theta(\alpha(u),\alpha(v))).
$$
Since $\alpha$ is an algebra morphism and
$\theta(\alpha(u),\alpha(v))=\theta(u,v)$, we deduce that
$\alpha_\eta$ is an algebra morphism.

By direct computations, we have
\begin{eqnarray*}
 && [\alpha_\eta(u,s),[(v,t),(w,m)]_\theta]_\theta+c.p.((u,s),(v,t),(w,m))\\
 &=&[(\alpha(u),s),([v,w]_\frkg,\theta(v,w))]_\theta+c.p.((u,s),(v,t),(w,m))\\
 &=&[\alpha(u),[v,w]_\frkg]_\frkg+c.p.(u,v,w)+\theta(\alpha(u),[v,w]_\frkg)+c.p.(u,v,w).
\end{eqnarray*}
Thus by the hom-Jacobi identity of $\frkg$, $[\cdot,\cdot]_\theta$
satisfies the hom-Jacobi identity if and only if
$$
\theta(\alpha(u),[v,w]_\frkg)+\theta(\alpha(v),[w,u]_\frkg)+\theta(\alpha(w),[u,v]_\frkg)=0,
$$
which exactly means that $\dM_T\theta=0.$ \qed

\begin{pro}
  For $\theta_1,\theta_2\in Z^2(\frkg)$, if $\theta_1-\theta_2$ is
  exact, the corresponding two central extensions
  $(\frkh,[\cdot,\cdot]_{\theta_1},\alpha_\frkh)$ and
  $(\frkh,[\cdot,\cdot]_{\theta_2},\alpha_\frkh)$ are isomorphic.
\end{pro}
\pf Assume that $\theta_1-\theta_2=\dM_Tf$, $f\in
C^1_\alpha(\frkg)$. Thus we have
$$
\theta_1(u,v)-\theta_2(u,v)=\dM_Tf(u,v)=-f([u,v]_\frkg).
$$
Define $f_\frkh:\frkh\longrightarrow\frkh$ by
$$
f_\frkh(u,s)=(u,s+f(u)).
$$
It is straightforward to see that
$f_\frkh\circ\alpha_\frkh=\alpha_\frkh\circ f_\frkh$. Obviously,
$f_\frkh$ is an isomorphism of vector spaces. We also have
\begin{eqnarray*}
  f_\frkh[(u,s),(v,t)]_{\theta_1}&=&f_\frkh([u,v]_\frkg,\theta_1(u,v))\\
  &=&([u,v]_\frkg,\theta_1(u,v)+f([u,v]_\frkg))\\
  &=&([u,v]_\frkg,\theta_2(u,v))\\
  &=&[f_\frkh(u,s),f_\frkh(v,t)]_{\theta_2}.
\end{eqnarray*}
Therefore, $f_\frkh$ is also an isomorphism of multiplicative
hom-Lie algebras. \qed

\section{The adjoint representation of hom-Lie algebras}

Let $(\frkg,\br,,\alpha)$ be a regular hom-Lie algebra. We consider
that $\frkg$ represents on itself via the bracket with respect to
the morphism $\alpha$. A very interesting phenomenon is that the
adjoint representation of  hom-Lie algebras is not unique as we will
see in the sequel.

\begin{defi}
 For any integer $s$, the $\alpha^s$-adjoint representation of the regular hom-Lie algebra
  $(\frkg,\br,,\alpha)$, which we denote by $\ad_s$, is defined by
  $$
\ad_s(u)(v)=[\alpha^s(u),v]_\frkg,\quad\forall~u,v\in\frkg.
  $$
\end{defi}
\begin{lem}
  With the above notations, we have
  \begin{eqnarray*}
    \ad_s(\alpha(u))\circ\alpha&=&\alpha\circ\ad_s(u);\\
\ad_s([u,v]_\frkg)\circ\alpha&=&\ad_s(\alpha(u))\circ\ad_s(v)-\ad_s(\alpha(v))\circ\ad_s(u).
  \end{eqnarray*}
  Thus the definition of $\alpha^s$-adjoint representation is well
  defined.
\end{lem}
\pf The conclusion follows from
\begin{eqnarray*}
    \ad_s(\alpha(u))(\alpha(v))&=&[\alpha^{s+1}(u),\alpha(v)]_\frkg\\
    &=&\alpha([\alpha^{s}(u),v]_\frkg)=\alpha\circ\ad_s(u)(v),
     \end{eqnarray*}
     and
\begin{eqnarray*}
\ad_s([u,v]_\frkg)(\alpha(w))&=&[\alpha^s([u,v]_\frkg),\alpha(w)]_\frkg\\
&=&[[\alpha^s(u),\alpha^s(v)]_\frkg,\alpha(w)]_\frkg\\
&=&[\alpha^{s+1}(u),[\alpha^s(v),w]_\frkg]_\frkg+[[\alpha^s(u),w]_\frkg,\alpha^{s+1}(v)]_\frkg\\
&=&\ad_s(\alpha(u))(\ad_s(v)(w))-\ad_s(\alpha(v))(\ad_s(u)(w)). \qed
  \end{eqnarray*}

 The set of
$k$-hom-cochains on $\frkg$ with coefficients in $\frkg$, which we
denote by $C^k_\alpha(\frkg;\frkg)$, is given by
$$
C^k_\alpha(\frkg;\frkg)=\{f\in
C^k(\frkg;\frkg)|~f\circ\alpha=\alpha\circ f\}.
$$
 In particular, the set of $0$-hom-cochains are given by:
$$C^0_\alpha(\frkg;\frkg)=\{u\in\frkg|\alpha(u)=u\}.$$

Associated to the $\alpha^{s}$-adjoint representation, the
coboundary operator $\dM_{s}:C^k_\alpha(\frkg;\frkg)\longrightarrow
C^{k+1}_\alpha(\frkg;\frkg)$ is given by
\begin{eqnarray*}
  \dM_{s} f(u_1,\cdots,u_{k+1})&=&\sum_{i=1}^{k+1}(-1)^{i+1}[\alpha^{k+s}(u_i),f(u_1,\cdots,\widehat{u_i},\cdots,u_{k+1})]\\
  &&+\sum_{i<j}(-1)^{i+j}f([u_i,u_j]_\frkg,\alpha(u_1)\cdots,\widehat{u_i},\cdots,\widehat{u_j},\cdots,\alpha(u_{k+1})).
\end{eqnarray*}

 For the $\alpha^s$-adjoint representation $\ad_s$, we obtain the
$\alpha^s$-adjoint complex $(C^\bullet_\alpha(\frkg;\frkg),\dM_s)$
and the corresponding cohomology
$$
H^k(\frkg;\ad_s)=Z^k(\frkg;\ad_s)/B^k(\frkg;\ad_s).
$$

\begin{rmk}
  In \cite{AEM}, the authors obtain a cochain complex for multiplicative hom-Lie algebra
  independently. It is straightforward to see that the cochain
  complex therein corresponds to $\alpha^{-1}$-adjoint
  representation. Here we focus on representations, thus we require
  that
  $\alpha$ is invertible, i.e. $\alpha^{-1}$ exists. As did in
  \cite{AEM}, if we do not care about representations, but only the
  cochain complex and the resulting cohomology $H^k(\frkg;\ad_s)~(k>0)$, the condition that
  $\alpha$ is invertible can be omitted.
\end{rmk}

In the  case of Lie algebras, a $1$-cocycle associated to the
adjoint representation is a  derivation. Similarly, we have

\begin{pro}\label{pro:derivation}
  Associated to the $\alpha^s$-adjoint representation $\ad_s$ of the
  regular
  hom-Lie algebra $(\frkg,\br,,\alpha)$, $D\in
  C^1_\alpha(\frkg;\frkg)$ is a $1$-cocycle if and only if $D$ is an
  $\alpha^{s+1}$-derivation, i.e. $D\in\Der_{\alpha^{s+1}}(\frkg)$.
\end{pro}
\pf The conclusion follows directly from the definition of the
coboundary operator $\dM_s$. $D$ is closed if and only if
$$
\dM_s(D)(u,v)=[\alpha^{s+1}(u),D(v)]_\frkg-[\alpha^{s+1}(v),D(u)]_\frkg-D([u,v]_\frkg)=0,
$$
which implies that $D$ is an
  $\alpha^{s+1}$-derivation. \qed

\subsection{The $\alpha^{-1}$-adjoint representation
$\ad_{-1}$}
\begin{pro}
  Associated to the  $\alpha^{-1}$-adjoint representation
$\ad_{-1}$, we have
\begin{eqnarray*}
H^0(\frkg;\ad_{-1})&=&\{u\in\frkg|\alpha(u)=u,~[u,v]_\frkg=0,\quad\forall~v\in\frkg\};\\
H^1(\frkg;\ad_{-1})&=&\Der_{\alpha^0}(\frkg)/\Inn_{\alpha^0}(\frkg).
\end{eqnarray*}
\end{pro}
\pf For any $0$-hom-cochain $u\in C^0_\alpha(\frkg;\frkg)$, we have
$$
\dM_{-1} u(v)=[\alpha^{-1}(v),u]_\frkg,\quad \forall~v\in\frkg.
$$
Therefore, $u$ is a closed $0$-hom-cochain if and only if
$[\alpha^{-1}(v),u]_\frkg=0$, which is equivalent to that
$\alpha([\alpha^{-1}(v),u]_\frkg)=[v,u]_\frkg=0$, which implies the
first conclusion. By Proposition \ref{pro:derivation}, we have $
Z^1(\frkg;\ad_{-1})=\Der_{\alpha^0}(\frkg).$ It is obvious that
$B^1(\frkg;\ad_{-1})=\Inn_{\alpha^0}(\frkg)$. Thus, we have
$H^1(\frkg;\ad_{-1})=\Der_{\alpha^0}(\frkg)/\Inn_{\alpha^0}(\frkg).$
\qed\vspace{3mm}

Let $\omega\in C^2_\alpha(\frkg;\frkg)$ be a skew-symmetric bilinear
operator commuting with $\alpha$. Consider a $t$-parametrized family
of bilinear operations
\begin{equation}
  [u,v]_t=[u,v]_\frkg+t\omega(u,v).
\end{equation}
Since $\omega$ commutes with $\alpha$, $\alpha$ is a morphism with
respect to the brackets $[\cdot,\cdot]_t$ for every $t$. If all the
brackets $[\cdot,\cdot]_t$ endow $(\frkg,[\cdot,\cdot]_t,\alpha)$
regular hom-Lie algebra structures, we say that $\omega$ generates a
deformation of the regular hom-Lie algebra
$(\frkg,[\cdot,\cdot]_\frkg,\alpha)$. By computing the hom-Jacobi
identity of $[\cdot,\cdot]_t$, this is equivalent to the conditions
\begin{eqnarray}\label{deformationC1}
  \omega(\alpha(u),[v,w]_\frkg)+[\alpha(u),\omega(v,w)]_\frkg+c.p.(u,v,w)&=&0;\\\label{deformationC2}
  \omega(\alpha(u),\omega(v,w))+c.p.(u,v,w)&=&0.
\end{eqnarray}
Obviously, \eqref{deformationC1} means that $\omega$ is closed with
respect to the $\alpha^{-1}$-adjoint representation $\ad_{-1}$, i.e.
$\dM_{-1}\omega=0.$ Furthermore, \eqref{deformationC2} means that
$\omega$ must itself define a hom-Lie algebra structure on $\frkg$.

A deformation is said to be trivial if there is a linear operator
$N\in C^1_\alpha(\frkg;\frkg)$ such that for $T_t={\Id}+tN$, there
holds
\begin{equation}
  T_t[u,v]_t=[T_t(u),T_t(v)]_\frkg.
\end{equation}
\begin{defi}
A linear operator $N\in C^1_\alpha(\frkg;\frkg)$ is called a
hom-Nijenhuis operator if we have
\begin{equation}\label{NijenhuisC}
  [Nu,Nv]_\frkg=N[u,v]_N,
\end{equation}
where the bracket $[\cdot,\cdot]_N$ is defined by
$$
[u,v]_N\triangleq[Nu,v]_\frkg+[u,Nv]_\frkg-N[u,v]_\frkg.
$$
\end{defi}

Please see \cite{Dorfman} for more details about Nijenhuis operators
of Lie algebras.

\begin{thm}
  Let $N\in C^1_\alpha(\frkg;\frkg)$ be a hom-Nijenhuis operator. Then a
  deformation of the regular hom-Lie algebra $(\frkg,\br,,\alpha)$ can be
  obtained by putting
  $$
\omega(u,v)=\dM_{-1}N(u,v)=[u,v]_N.
  $$
  Furthermore, this deformation is trivial.
\end{thm}
\pf Since $\omega=\dM_{-1}N$, $\dM_{-1}\omega=0$ is valid. To see
that $\omega$ generates a deformation, we need to check the
hom-Jacobi identity for $\omega$. Using the explicit expression of
$\omega$, we have
\begin{eqnarray*}
  &&\omega(\alpha(u),\omega(v,w))+c.p.(u,v,w)\\
  &=&[N\alpha(u),[Nv,w]_\frkg]_\frkg+[N\alpha(v),[w,Nu]_\frkg]_\frkg+[\alpha(w),N[u,v]_N]_\frkg+c.p.(u,v,w)\\
  &&+N[\alpha(v),N[w,u]_\frkg]_\frkg-[N\alpha(v),N[w,u]_\frkg]_\frkg+c.p.(u,v,w)\\
  &&-N\big([\alpha(u),[Nv,w]_\frkg]_\frkg+[\alpha(w),[u,Nv]_\frkg]_\frkg\big)+c.p.(u,v,w).
\end{eqnarray*}
Since $N$ commutes with $\alpha$, by the hom-Jacobi identity of
$\frkg$, we have
$$
[N\alpha(u),[Nv,w]_\frkg]_\frkg+[N\alpha(v),[w,Nu]_\frkg]_\frkg=[[Nu,Nv]_\frkg,\alpha(w)]_\frkg.
$$
Since $N$ is a hom-Nijenhuis operator, we have
$$
[N\alpha(u),[Nv,w]_\frkg]_\frkg+[N\alpha(v),[w,Nu]_\frkg]_\frkg+[\alpha(w),N[u,v]_N]_\frkg+c.p.(u,v,w)=0.
$$
Furthermore, also by the fact that $N$ is a hom-Nijenhuis operator,
we have
$$
N[\alpha(v),N[w,u]_\frkg]_\frkg-[N\alpha(v),N[w,u]_\frkg]_\frkg=-N[N\alpha(v),[w,u]_\frkg]_\frkg+N^2[\alpha(v),[w,u]_\frkg]_\frkg.
$$
Thus by the hom-Jacobi identity of $\frkg$, we have
$$
N[\alpha(v),N[w,u]_\frkg]_\frkg-[N\alpha(v),N[w,u]_\frkg]_\frkg+c.p.(u,v,w)=-N[N\alpha(v),[w,u]_\frkg]_\frkg+c.p.(u,v,w).
$$
Therefore, we have
\begin{eqnarray*}
  &&\omega(\alpha(u),\omega(v,w))+c.p.(u,v,w)\\
  &=&-N[N\alpha(v),[w,u]_\frkg]_\frkg-N\big([\alpha(u),[Nv,w]_\frkg]_\frkg+[\alpha(w),[u,Nv]_\frkg]_\frkg\big)+c.p.(u,v,w)\\
  &=&-N\big([\alpha(Nv),[w,u]_\frkg]_\frkg+[\alpha(u),[Nv,w]_\frkg]_\frkg+[\alpha(w),[u,Nv]_\frkg]_\frkg\big)+c.p.(u,v,w)\\
  &=&0.
\end{eqnarray*}
Thus $\omega$ generates a deformation of the hom-Lie algebra
$(\frkg,\br,,\alpha)$.

Let $T_t={\Id} +tN$, then we have
\begin{eqnarray*}
  T_t[u,v]_t&=&({\Id}+tN)([u,v]_\frkg+t[u,v]_N)\\
  &=&[u,v]_\frkg+t([u,v]_N+N[u,v]_\frkg)+t^2N[u,v]_N.
\end{eqnarray*}
On the other hand, we have
\begin{eqnarray*}
  [T_t(u),T_t(v)]_\frkg&=&[u+tNu,v+tNv]_\frkg\\
  &=&[u,v]_\frkg+t([Nu,v]_\frkg+[u,Nv]_\frkg)+t^2[Nu,Nv]_\frkg.
\end{eqnarray*}
By  \eqref{NijenhuisC}, we have
$$
T_t[u,v]_t=[T_t(u),T_t(v)]_\frkg,
$$
which implies that the deformation is trivial. \qed

\subsection{The $\alpha^{0}$-adjoint representation
$\ad_{0}$}
\begin{pro}
Associated to the $\alpha^{0}$-adjoint representation $\ad_{0}$, we
have\begin{eqnarray*}
H^0(\frkg;\ad_{0})&=&\{u\in\frkg|\alpha(u)=u,~[u,v]_\frkg=0,~\forall~v\in \frkg\};\\
H^1(\frkg;\ad_{0})&=&\Der_{\alpha}(\frkg)/\Inn_{\alpha}(\frkg).
\end{eqnarray*}
\end{pro}
\pf For any $0$-hom-cochain $u\in C^0_\alpha(\frkg;\frkg)$, we have
$$
\dM_0 u(v)=[\alpha^{0}(v),u]_\frkg=[v,u]_\frkg,\quad
\forall~v\in\frkg.
$$
Therefore, the set of closed  $0$-hom-cochain $Z^0(\frkg;\ad_0)$ is
given by
$$
Z^0(\frkg;\ad_0)=\{u\in
C^0_\alpha(\frkg;\frkg)|[u,v]_\frkg=0,~\forall~v\in \frkg\}.
$$
 Thus we have
$$
H^0(\frkg;\ad_0)=\{u\in\frkg|\alpha(u)=u,~[u,v]_\frkg=0,~\forall~v\in
\frkg\}.
$$
For any $f\in C^1_\alpha(\frkg;\frkg)$, we have
$$
\dM_{0}f(u,v)=[\alpha(u),f(v)]_\frkg-[\alpha(v),f(u)]_\frkg-f([u,v]_\frkg).
$$
Therefore, the set of closed $1$-hom-cochain $Z^1(\frkg;\ad_0)$ is
exactly the set of $\alpha$-derivation $\Der_{\alpha}(\frkg)$.
Furthermore, it is obvious that  any exact $1$-hom-cochain is of the
form $[\cdot,u]_\frkg$ for some $u\in C^0_\alpha(\frkg;\frkg)$.
Therefore, we have $B^1(\frkg;\ad_0)=\Inn_{\alpha}(\frkg)$, which
implies that $
H^1(\frkg;\ad_0)=\Der_{\alpha}(\frkg)/\Inn_{\alpha}(\frkg).
$\qed\vspace{3mm}

In Section 3 we have already proved that associated to any
$\alpha$-derivation $D$ of the hom-Lie algebra
$(\frkg,\br,,\alpha)$, there is a derivation extension
$(\frkg\oplus\mathbb RD,[\cdot,\cdot]_D,\alpha_D)$. Thus the
derivation extension of the hom-Lie algebra $(\frkg,\br,,\alpha)$ is
controlled by its first cohomology with coefficients in the
$\alpha^0$-adjoint representation.

\section*{Appendix: The proof of Proposition \ref{pro:coboundary}}

 By straightforward computations, we have
\begin{eqnarray} \nonumber &&
\dM_{\rho_A}^2f(u_1,\cdots,u_{k+2})\\\nonumber
 &=&\sum_{i=1}^{k+2}(-1)^{i+1}\rho_A(\alpha^{k+1}(u_i))(\dM_{\rho_A}
 f(u_1,\cdots,\widehat{u_i},\cdots,u_{k+2}))\\\nonumber
 &&+\sum_{i<j}(-1)^{i+j}\dM_{\rho_A}
 f([u_i,u_j]_\frkg,\alpha(u_1),\cdots,\widehat{u_i},\cdots,\widehat{u_j},\cdots,\alpha(u_{k+2})).
\end{eqnarray}
It is not hard to deduce that

\begin{eqnarray}&&\nonumber\rho_A(\alpha^{k+1}(u_i))(\dM_{\rho_A}
 f(u_1,\cdots,\widehat{u_i},\cdots,u_{k+2})) \\\label{t1}
&=&\sum_{p=1}^{i-1}(-1)^{p+1}\rho_A(\alpha^{k+1}(u_i))\circ\rho_A(\alpha^{k}(u_p))(f(u_1,\cdots,\widehat{u_p},\cdots,\widehat{u_i},\cdots,u_{k+2}))\\\label{t2}
&&
+\sum_{p=i+1}^{k+2}(-1)^p\rho_A(\alpha^{k+1}(u_i))\circ\rho_A(\alpha^{k}(u_p))(f(u_1,\cdots,\widehat{u_i},\cdots,\widehat{u_p},\cdots,u_{k+2}))\\\nonumber
&&
+\sum_{i<p<q}(-1)^{p+q}\rho_A(\alpha^{k+1}(u_i))(f([u_p,u_q]_\frkg,\alpha(u_1),\cdots,\widehat{u_i},\cdots,\widehat{u_p},\cdots\widehat{u_q}\cdots,\alpha(u_{k+2})))\\\nonumber
&&
+\sum_{p<q<i}(-1)^{p+q}\rho_A(\alpha^{k+1}(u_i))(f([u_p,u_q]_\frkg,\alpha(u_1),\cdots,\widehat{u_p},\cdots,\widehat{u_q},\cdots,\widehat{u_i},\cdots,\alpha(u_{k+2})))\\\nonumber
&&
+\sum_{p<i<q}(-1)^{p+q+1}\rho_A(\alpha^{k+1}(u_i))(f([u_p,u_q]_\frkg,\alpha(u_1),\cdots,\widehat{u_p},\cdots,\widehat{u_i},\cdots,\widehat{u_q}\cdots,\alpha(u_{k+2}))),
\end{eqnarray}
and
\begin{eqnarray}
&&\nonumber\dM_{\rho_A}
 f([u_i,u_j]_\frkg,\alpha(u_1),\cdots,\widehat{u_i},\cdots,\widehat{u_j},\cdots,\alpha(u_{k+2}))\\\label{t3}
 &=&
 \rho_A(\alpha^{k}[u_i,u_j]_\frkg)(
 f(\alpha(u_1),\cdots,\widehat{u_i},\cdots,\widehat{u_j},\cdots,\alpha(u_{k+2})))\\\nonumber
 &&+\sum_{p=1}^{i-1}(-1)^p
\rho_A(\alpha^{k+1}(u_p))(
 f([u_i,u_j]_\frkg,\alpha(u_1),\cdots,\widehat{u_p},\cdots\widehat{u_i},\cdots,\widehat{u_j},\cdots,\alpha(u_{k+2})))\\\nonumber
 &&+\sum_{p=i+1}^{j-1}(-1)^{p+1}
 \rho_A(\alpha^{k+1}(u_p))(
 f([u_i,u_j]_\frkg,\alpha(u_1),\cdots,\widehat{u_i},\cdots,\widehat{u_p},\cdots,\widehat{u_j},\cdots,\alpha(u_{k+2})))\\\nonumber
 &&+\sum_{p=j+1}^{k+2}(-1)^{p}
 \rho_A(\alpha^{k+1}(u_p))(
 f([u_i,u_j]_\frkg,\alpha(u_1),\cdots,\widehat{u_i},\cdots,\widehat{u_j},\cdots,\widehat{u_p},\cdots,\alpha(u_{k+2})))\\\label{t4}
 &&+\sum_{p=1}^{i-1}(-1)^{1+p}
 f([[u_i,u_j]_\frkg,\alpha(u_p)]_\frkg,\alpha^2(u_1),\cdots,\widehat{u_p},\cdots,\widehat{u_i},\cdots,\widehat{u_j},\cdots,\alpha^2(u_{k+2}))\\\label{t5}
 &&+\sum_{p=i+1}^{j-1}(-1)^{p}
 f([[u_i,u_j]_\frkg,\alpha(u_p)]_\frkg,\alpha^2(u_1),\cdots,\widehat{u_i},\cdots,\widehat{u_p},\cdots,\widehat{u_j},\cdots,\alpha^2(u_{k+2}))\\\label{t6}
 &&+\sum_{p=j+1}^{k+2}(-1)^{p+1}
 f([[u_i,u_j]_\frkg,\alpha(u_p)]_\frkg,\alpha^2(u_1),\cdots,\widehat{u_i},\cdots,\widehat{u_j},\cdots,\widehat{u_p},\cdots,\alpha^2(u_{k+2}))\\\label{t7}
&&+\sum_{p,q}(\pm)
 f([\alpha(u_p),\alpha(u_q)]_\frkg,\alpha[u_i,u_j]_\frkg,\alpha^2(u_1),\cdots,\widehat{u_{i,j,p,q}},\cdots,\alpha^2(u_{k+2})).
\end{eqnarray}

In \eqref{t7}, $\widehat{u_{i,j,p,q}}$ means that we omit the items
$u_i,u_j,u_p,u_q$. By the fact that $\alpha$ is an algebra morphism,
i.e. $[\alpha(u_p),\alpha(u_q)]_\frkg=\alpha[u_p,u_q]_\frkg$, we get
$$\sum_{i<j}(-1)^{i+j}\eqref{t7}=0.$$
By the hom-Jacobi identity, we obtain that
$$\sum_{i<j}(-1)^{i+j}\big(\eqref{t4}+\eqref{t5}+\eqref{t6}\big)=0.$$
At last, note that $f$ commutes with $\alpha$, we have that
\eqref{t3} equals to
$$
 \rho_A([\alpha^{k}(u_i),\alpha^{k}(u_j)]_\frkg)\circ
A\circ
 f(u_1,\cdots,\widehat{u_i},\cdots,\widehat{u_j},\cdots,u_{k+2}).
$$
Thus by \eqref{representation2}, we have
$$
\sum_{i=1}^{k+2}(-1)^{i+1}\big(\eqref{t1}+\eqref{t2}\big)+\sum_{i<j}(-1)^{i+j}\eqref{t3}=0.
$$
The sum of the other six items is zero obviously. Therefore, we have
$\dM_{\rho_A}^2=0$. The proof is completed.


\begin{thebibliography}{999}


\bibitem{AEM} F. Ammar, Z. Ejbehi and A. Makhlouf, Cohomology and Deformations
of Hom-algebras, arXiv:1005.0456.

\bibitem{BM}
S. Benayadi and A. Makhlouf, Hom-Lie Algebras with Symmetric
Invariant NonDegenerate Bilinear Forms, arXiv:1009.4226.

\bibitem{CE}
C. Chevalley and S. Eilenberg, Cohomology theory of Lie groups and
Lie algebras, \emph{ Trans. Amer. Math. Soc.} 63 (1948) 85-124.

\bibitem{Dorfman}  I. Dorfman,  {\em Dirac Structures and
    Integrability of Nonlinear Evolution Equation.} Wiley, 1993.

\bibitem{HLS}
J. Hartwig, D. Larsson and S. Silvestrov, Deformations of Lie
algebras using $\sigma$-derivations, \emph{J. Algebra} 295 (2006),
314-361.

\bibitem{Jacobson} N. Jacobson,  {\em Lie algbras}. Dover Publications, Inc. New
York (1962).

\bibitem{LD1}
D. Larsson and S. Silvestrov,  Quasi-hom-Lie algebras, central
extensions and 2-cocycle-like identities, \emph{ J. Algebra} 288
(2005) 321-344.

\bibitem{LD2}
D. Larsson and S. Silvestrov, Quasi-Lie algebras, \emph{Contemp.
Math.} 391 (2005) 241-248.



\bibitem{MS1}
 A. Makhlouf and S. Silvestrov, Notes on Formal Deformations of Hom-associative and Hom-Lie
 Algebras, \emph{Forum Math.}, 22 (2010), no. 4, 715-739.

\bibitem{MS2} A. Makhlouf and S. Silvestrov, Hom-algebra
structures, \emph{J. Gen. Lie Theory Appl.} Vol. 2 (2008), No. 2,
51-64.



\bibitem{sheng}
Y. Sheng, Linear Poisson structures on $\mathbb R^4$, \emph{ J.
Geom. Phys.} 57 (2007), 2398-2410.

\bibitem{Yao1}
 D. Yau, Hom-Yang-Baxter equation, Hom-Lie algebras, and quasi-triangular
 bialgebras, \emph{J. Phys. A: Math. Theor.} 42 (2009), 165202.

\bibitem{Yao2}
D. Yau, Hom-algebras and homology, \emph{ J. Lie Theory} 19 (2009)
409-421.

\bibitem{Yao3}
D. Yau, Enveloping algebras of Hom-Lie algebras, \emph{J. Gen. Lie
Theory Appl.} 2 (2008), 95-108.

\end{thebibliography}
\end{document}